\documentclass[11pt,preprint]{aastex}

\def\gax{\mathrel{\raise.3ex\hbox{$>$}\mkern-14mu\lower0.6ex\hbox{$\sim$}}}
\def\lax{\mathrel{\raise.3ex\hbox{$<$}\mkern-14mu\lower0.6ex\hbox{$\sim$}}}
\def\gtorder{\mathrel{\raise.3ex\hbox{$>$}\mkern-14mu
             \lower0.6ex\hbox{$\sim$}}}
\def\ltorder{\mathrel{\raise.3ex\hbox{$<$}\mkern-14mu
             \lower0.6ex\hbox{$\sim$}}}

\begin{document}

\title{A Simple Method To Find All Lensed Quasars\footnote{Based on observations
obtained with the 1.3m telescope of the Small and Moderate Aperture
Research Telescope System (SMARTS), which is operated by the SMARTS
Consortium.}} 

\author{C.S. Kochanek, B. Mochejska, N.D. Morgan \& K.Z. Stanek}
 
\affil{\small Department of Astronomy, The Ohio State University, 140
West 18th Avenue, Columbus OH 43210 \\
  ckochanek, bmochejs, nmorgan, kstanek@astronomy.ohio-state.edu}

\begin{abstract}
We demonstrate that gravitationally lensed quasars are easily recognized 
using image subtraction methods as time variable sources that are spatially extended.  For Galactic
latitudes $|b|\gtorder 20^\circ$, lensed quasars dominate the population of
spatially extended variable sources, although there is some contamination
from variable star pairs, variable star-quasar pairs and binary quasars
that can be easily controlled using other information in the survey
such as the object light curves and colors.  This will allow
planned large-scale synoptic surveys to find lensed quasars
almost down to their detection limits without the need for extensive
follow-up observations.
\end{abstract}


\section{Introduction and Method}
\label{sec:introduction}

In theory, gravitational lenses can be used to address astrophysical problems
such as the cosmological model, the structure and evolution of galaxies, 
and the structure of quasar accretion disks (see the reviews by 
Kochanek~(\citeyear{Kochanek2004saas}) of strong lensing and Wambsganss
(\citeyear{Wambsganss2004saas}) of microlensing).
One of the main challenges in using lenses for any of these applications is 
discovering large numbers of lenses efficiently (see the review of lens
surveys in Kochanek~(\citeyear{Kochanek2004saas})).  Most known lenses have
been found either from optical imaging surveys of known quasars 
(see  Pindor et al.~\citeyear{Pindor2003p2340} for a recent study),
radio imaging surveys of flat-spectrum radio sources 
(see Browne et al.~\citeyear{Browne2003p13}), or searches for 
anomalous, higher redshift emission lines in galaxy spectra
(see Bolton et al.~\citeyear{Bolton2005}).  Imaging surveys of all radio
sources (Burke~\citeyear{Burke1990}) proved difficult because of the
confusing array of structures observed for steep spectrum radio
sources.  Haarsma et al.~(\citeyear{Haarsma2005}) proposed improving
the efficiency of searches for lensed steep-spectrum sources by
looking for radio lobes with optical counterparts, but the approach
is limited by the resolution and sensitivity of existing all-sky radio surveys.

None of these methods is easily applied to the next generation of
large scale imaging surveys such as the SDSS Supernova Survey
(Sako et al.~\citeyear{Sako2005}), the Dark Energy Survey (DES, Abbott et al.~\citeyear{Abbott2005}),
Pan-STARRS (Kaiser~\citeyear{Kaiser2004}) and the Large Synoptic Survey Telescope
(LSST, Tyson et al.~\citeyear{Tyson2003}).    One possibility is to 
use a combination of color and morphology to identify quasar lens candidates 
(Morgan et al.~\citeyear{Morgan2004}).  This strategy can be effective as 
long as emission (or absorption) by the lens galaxy does not significantly
change the color of the system from that of the quasars, which restricts
its applicability to systems in which the quasar images are significantly
brighter than the lens galaxy.  A new feature of all these
projects, however, is that they are synoptic surveys which obtain
light curves for variable sources.  Pindor (\citeyear{Pindor2005p649}) suggested 
that the synoptic data could be used to find lenses by cross-correlating the
light curves of closely separated sources to search for the time delays 
present in the lensed systems.  This approach may be problematic as a 
search method because it requires the automated extraction of light
curves for the individual lensed images, some of which may also be
distorted by the effects of microlensing.  However, it will be an essential 
component of verifying lens candidates in the synoptic surveys.

In this paper we introduce a far simpler strategy.  Unlike almost any other
source, lensed quasars are ``extended'' variable sources because the variable
flux is spread out over the scale of the image separations.  As we discuss
in \S2, restricting the search to extended variable sources is an extraordinarily 
powerful means of eliminating sources other than gravitational lenses.
In \S3 we demonstrate the method using data we have been acquiring to 
measure time delays and microlensing variability in known lensed quasars
(Kochanek et al.~\citeyear{Kochanek2005}).  We summarize our proposed search in \S4.

\section{Variability As A Means of Eliminating False Positives}

\begin{figure}
\plotone{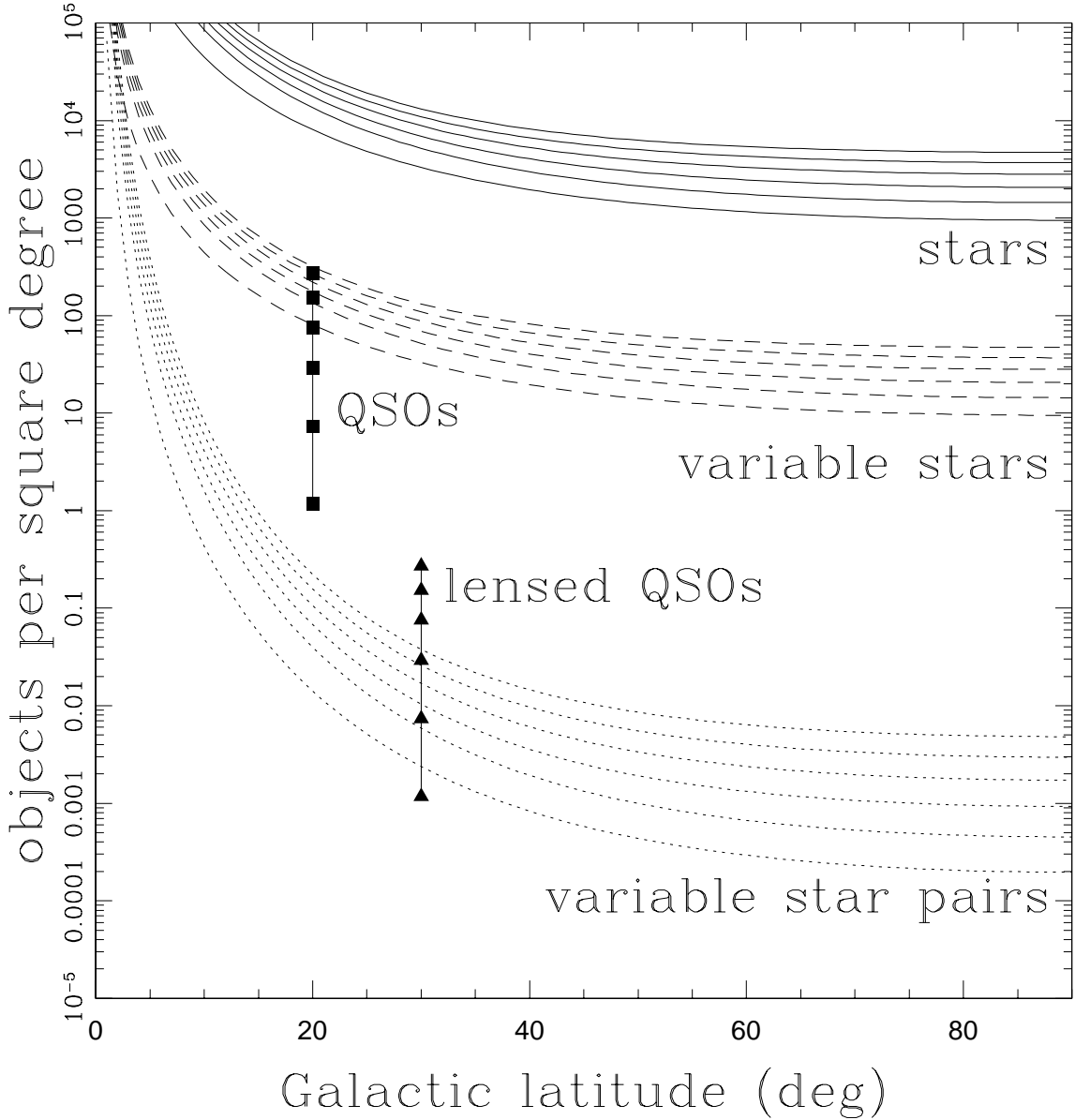}
\caption{ Background suppression from detecting gravitational lenses as 
  extended variable sources.  The solid curves (top, labeled ``stars'')
  show the integral surface density of 19-24~V mag stars as a function
  of Galactic latitude (at $\ell=90^\circ$ from the Bahcall \& Soniera  
  (\citeyear{Bahcall1980p17}) model) in steps of 1~mag.  The dashed curves (middle, labeled
  ``variable stars'') show the surface density of variable stars assuming
  $f_{*,var}=0.01$ of stars are variables.  The dotted curves 
   (bottom, labeled ``variable star pairs'') shows the surface density
  of pairs of variable stars with separations $\Delta\theta < 3\farcs0$.
  For comparison, the filled squares show the surface density of
  $1 < z < 2.5$ quasars with 19-24~g mag based on the 2SLAC luminosity
  function, and the filled triangles show the surface density of
  variable lensed quasars assuming that $f_{q,var} p_{lens} =10^{-3}$.
  For $|b| \gtorder 20^\circ$, lensed quasars are more common
  than variable star pairs.
  }
\label{fig:starcount} 
\end{figure}

The basic problem in lens searches is that they are intrinsically rare objects.
We start with the problem that quasars are relatively rare.   Fig.~\ref{fig:starcount} 
shows the surface density of quasars ($1 < z < 2.5$) computed from the g-band 2SLAQ quasar
luminosity functions (Richards et al.~\citeyear{Richards2005p839}).  For these models,
the surface density at 23~mag is approximately $10^2$~deg$^{-2}$.  Lensed quasars
are rarer still, since a conservative estimate for the lensing probability of these
faint quasars is $p_{lens}\simeq 0.002$ (see the review of lens statistics in
Kochanek~\citeyear{Kochanek2004saas}).   Thus, while the number of faint, lensed
quasars is almost two orders of magnitude greater than the number of lenses
presently known, it is not a trivial problem to find the one lensed quasar in
each 5~deg$^2$ region given the $\sim 2 \times 10^5$  other sources in the same
area.  The problem is further complicated by the increasing importance of the
contribution of the lens galaxy flux to the total flux of the lens as we search
for fainter lensed sources.  The lens galaxy masks both the color and morphology
of the lensed images, making traditional quasar selection methods useless.

The key to our approach is to apply difference imaging (Alard \& Lupton~\citeyear{Alard1998p325},
Alard~\citeyear{Alard2000p363}) to the synoptic data from large imaging surveys. Some
version of difference imaging will be used in all these surveys as the basis for 
identifying variable sources and extracting light curves.    Difference imaging
works by scaling, in both flux and PSF substructure, a reference image to match
the data obtained for each epoch and then subtracting the two to form a series
of difference images $R_i$.  The difference image has flux only for objects that have 
varied  between the reference image and the epoch under consideration,
so it has the immediate advantage of eliminating all the galaxies.  
We focus on time variability because quasars are intrinsically variable sources.
On two year time scales, roughly 50\% of quasars vary by more than 0.1~mag
(e.g. Cimatti et al.~\citeyear{Cimatti1993}) with general correlations that fainter 
quasars observed at bluer wavelengths show greater variability (Vanden Berk et al.~\citeyear{vandenberk2004}).  
The variability of lensed quasars will be still greater than that of unlensed 
quasars because they are also microlensed by the stars in the lens galaxy (see
Wambsganss~\citeyear{Wambsganss2004saas}).  We will conservatively assume that
fraction $f_{q,var} \simeq 0.5$ of detected quasars will show 10\% flux variations 
during the course of the survey.

We can divide variable sources into three general categories: variable
point sources (stars, quasars, supernovae and other explosive events), moving
solar system objects (asteroids, Kuiper belt objects), and gravitational lenses.
Let us consider what these three classes of objects look like in an image of the 
variable flux formed by computing the absolute value or root-mean-square (rms) of the
differenced images for a field.  The point sources are variable but do not move,
so they will appear as point sources.  Solar system objects move rapidly across
the field, appearing as a sparsely realized ``track.''  Small separation gravitationally
lensed quasars appear as extended, non-circular objects, while wider separation
lensed quasars appear as very closely separated groupings of variable objects.  We
will refer to them as examples of ``extended'' variable objects.  

While resolved four-image lenses are virtually impossible to mimic with any other
source, two-image lenses can be mimicked by several other sources.  The least
interesting backgrounds are variable star pairs or variable star/quasar pairs
which occur simply because of chance superpositions.  Two astrophysically
interesting backgrounds are binary quasars, whose abundance on 
separations $\Delta\theta\ltorder 10\farcs0$ is comparable to that of
gravitational lenses (see Hennawi et al~\citeyear{Hennawi2005}), and lensed
supernovae.  Very crudely, if there is one detectable supernovae per $<23$~mag 
galaxy per century, then the abundance of lensed supernovae is comparable to that
of lensed quasars.  Both binary quasars and lensed supernovae can be easily 
distinguished from lensed quasars based on their light curves.  This can be
done rapidly for the case of supernovae, but may take $\sim$years for the case
of a binary quasar. 

We focus on comparing the surface density of lenses to the surface density of
variable star pairs, since they represent the least interesting background.
Star pairs that can be resolved in normal seeing are separated by large 
physical distances and should show no significant correlations in their
activity.  This makes it straight forward to estimate
the background of uninteresting extended variable sources.     

While in general the fraction of stars that are variable will vary 
greatly depending on the stellar population observed, for a ``normal" 
stellar mix about 1-2\% of stars vary by more than 1\% (e.g. Hartman et 
al.~\citeyear{Hartman2004}). Moreover, studies of variable quasars located 
behind the SMC showed that their main stellar contaminant in the 
color-variability space are massive Be stars, which will be rare 
away from the Galactic plane (Dobrzycki et al.~\citeyear{Dobrzycki2003}).
Since we will be interested in regions well away from the Galactic
plane and the significantly higher long term variability amplitudes
of quasars, we will assume that fraction $f_{*,var} \sim 1\%$ of stars
will have variability comparable to that of quasars.  Thus, if
there are $\Sigma_* \simeq 4 \times 10^3$~deg$^{-2}$ stars 
with V$<23$~mag at high Galactic latitude (Bahcall \& Soneira~\citeyear{Bahcall1980p17})
the surface density of variable stars is $N_{var}=f_{*,var}\Sigma_* \simeq 40$~deg$^{-2}$. 
The surface density of (uncorrelated) pairs of variable stars separated 
by $\Delta\theta$ should be of order 
\begin{equation}
     \Sigma_{pair}\simeq \pi \Delta\theta^2 (f_{*,var}\Sigma_*)^2\simeq 
      0.0002 \left( {\Sigma_* \over 10^3\hbox{deg}^{-2}}
              { f_{*,var} \over 0.01} {\Delta\theta \over 3\farcs0 }\right)^2 \hbox{deg}^{-2}.
\end{equation}
for a surface density 1/300~deg$^2$ at V$<23$ with $\Delta\theta < 3\farcs0$ that is
well below that of gravitational lenses.
Fig.~\ref{fig:starcount} shows the expected surface density of stars, variable stars
and variable stars pairs ($\Delta\theta < 3\farcs0$) as a function of Galactic latitude 
(at $\ell=90^\circ$) and V-band magnitude in the Bahcall \& Soneira~(\citeyear{Bahcall1980p17}) 
star count models.  The surface density of star pairs  
with $\Delta\theta < 3\farcs0$ is comparable to the surface density of variable stars.
Variable star pairs and variable star/quasar pairs are comparable in abundance, with
the former being being more common for V$\ltorder 21$~mag and the latter more common
at fainter magnitudes.  
Comparing the surface density of variable star pairs and lensed quasars, we see that
samples of extended variable objects should be dominated by gravitational lenses rather
than chance superpositions.

\begin{figure}
\plotone{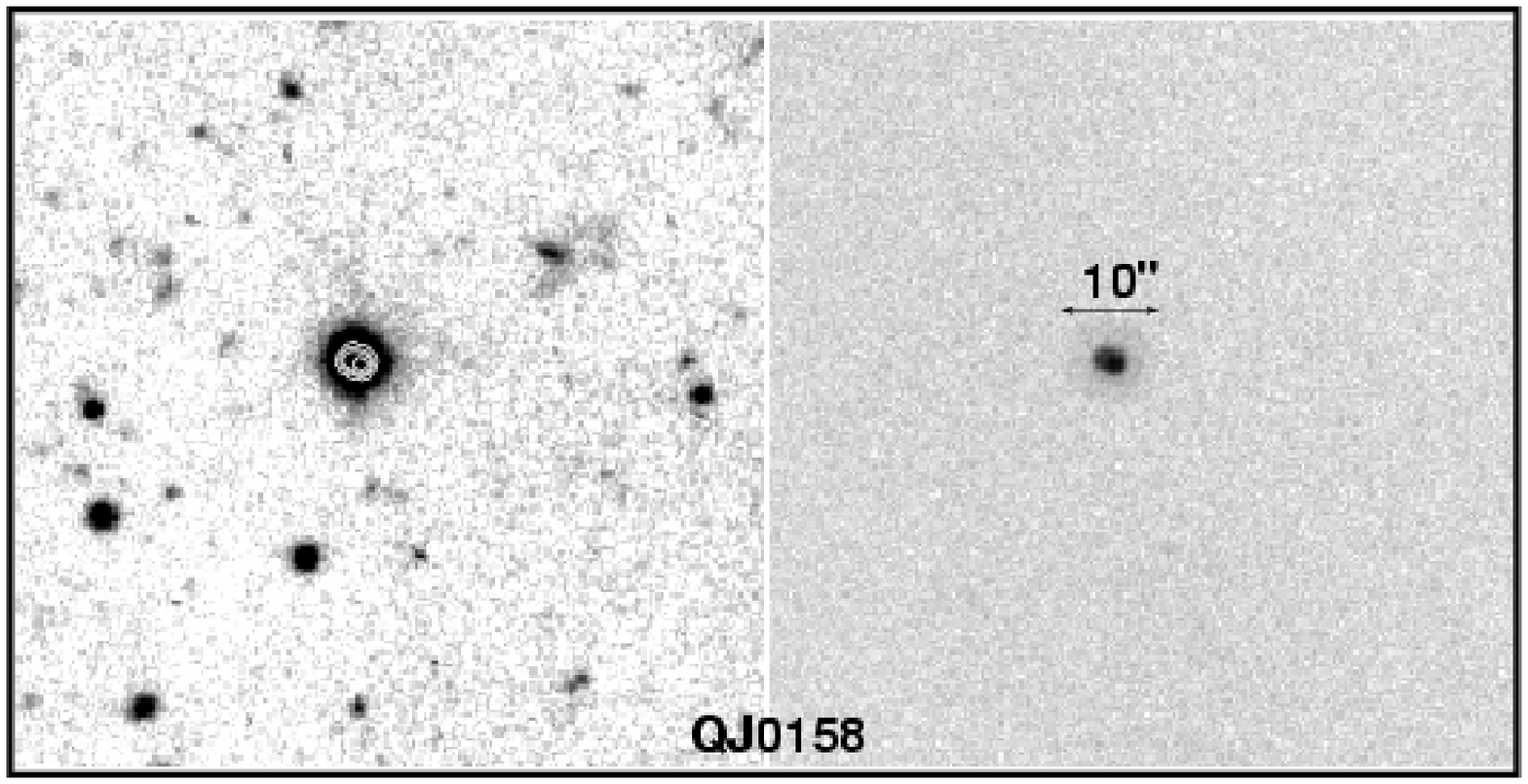}
\plotone{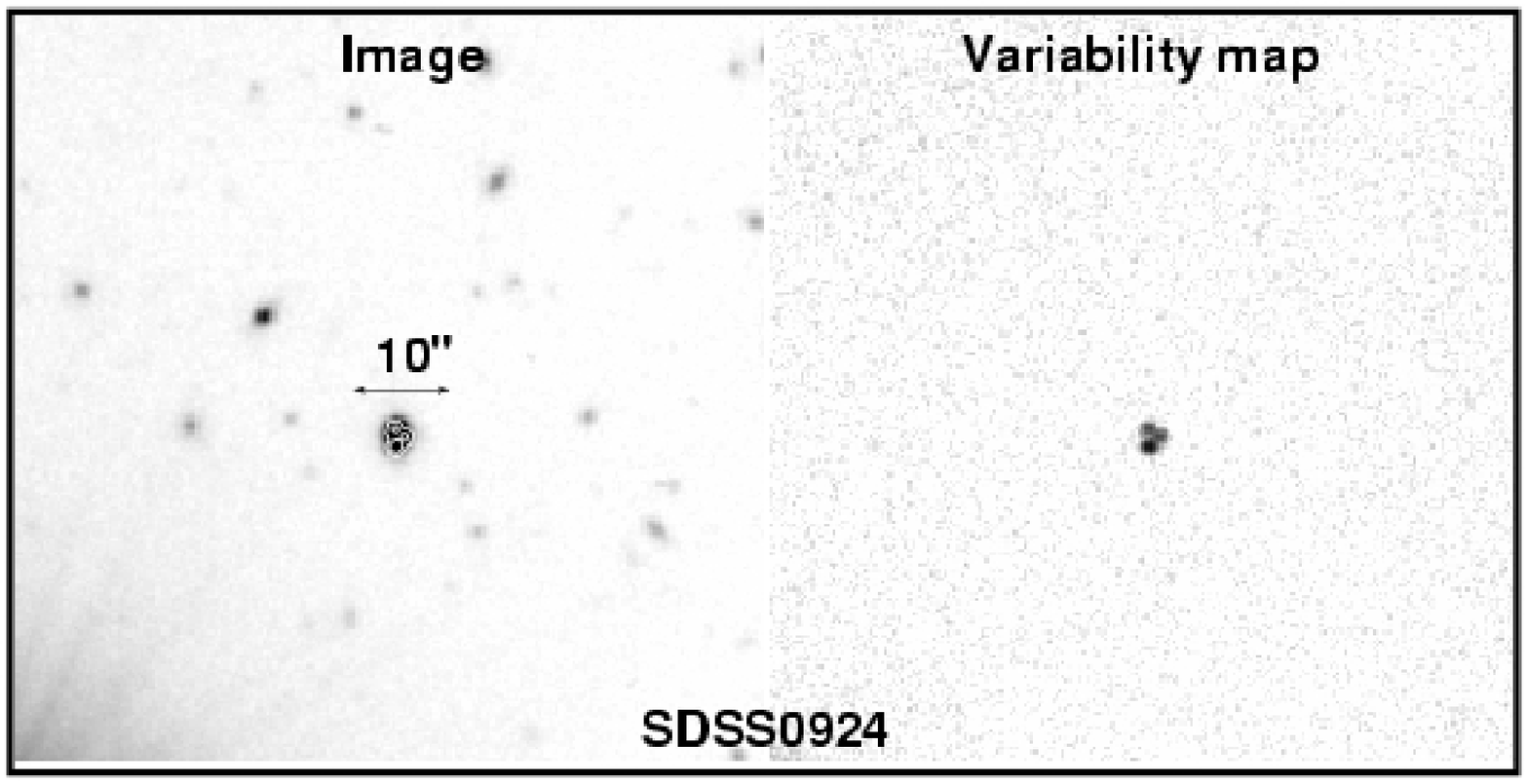}
\caption{
  Average ($I_T$, left) and variability significance ($R_R/I_N$, right) images of QJ~0158--4325 (top)
  and SDSS~0924+0219 (bottom).  The Einstein radii of the two lenses are 1\farcs2 and 1\farcs7,
  respectively.  The contours in the left panels are logarithmically 
  spaced contours of the variability image in the right panel shown to highlight the
  lens structure.  Note that all the objects other than
  the lens in the average image have vanished from the variability image and that
  the structures of the lenses in the variability significance images are
  easily distinguished from those of an unresolved point source.
  }
\label{fig:lenses1}
\end{figure}

\begin{figure}
\plotone{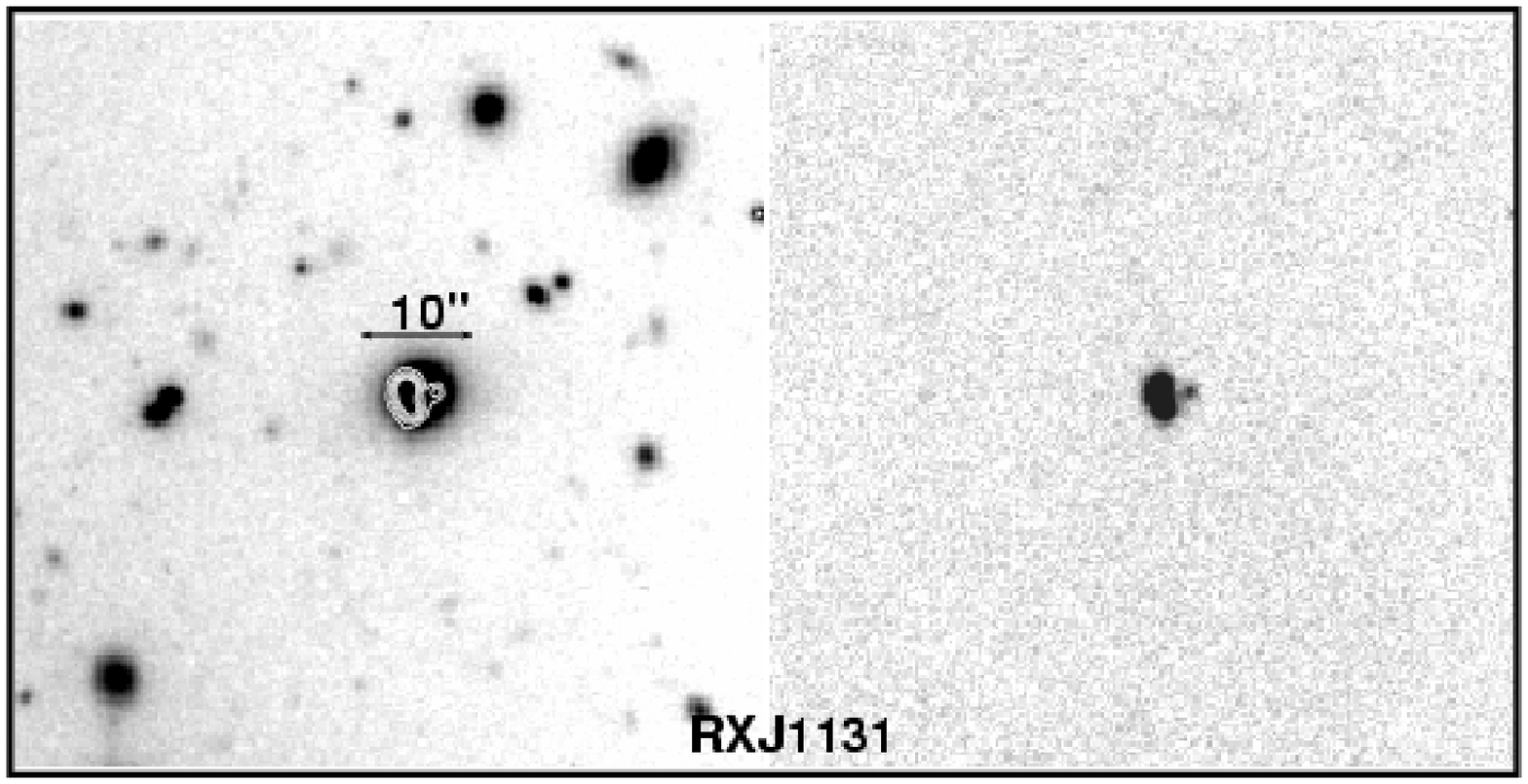}
\plotone{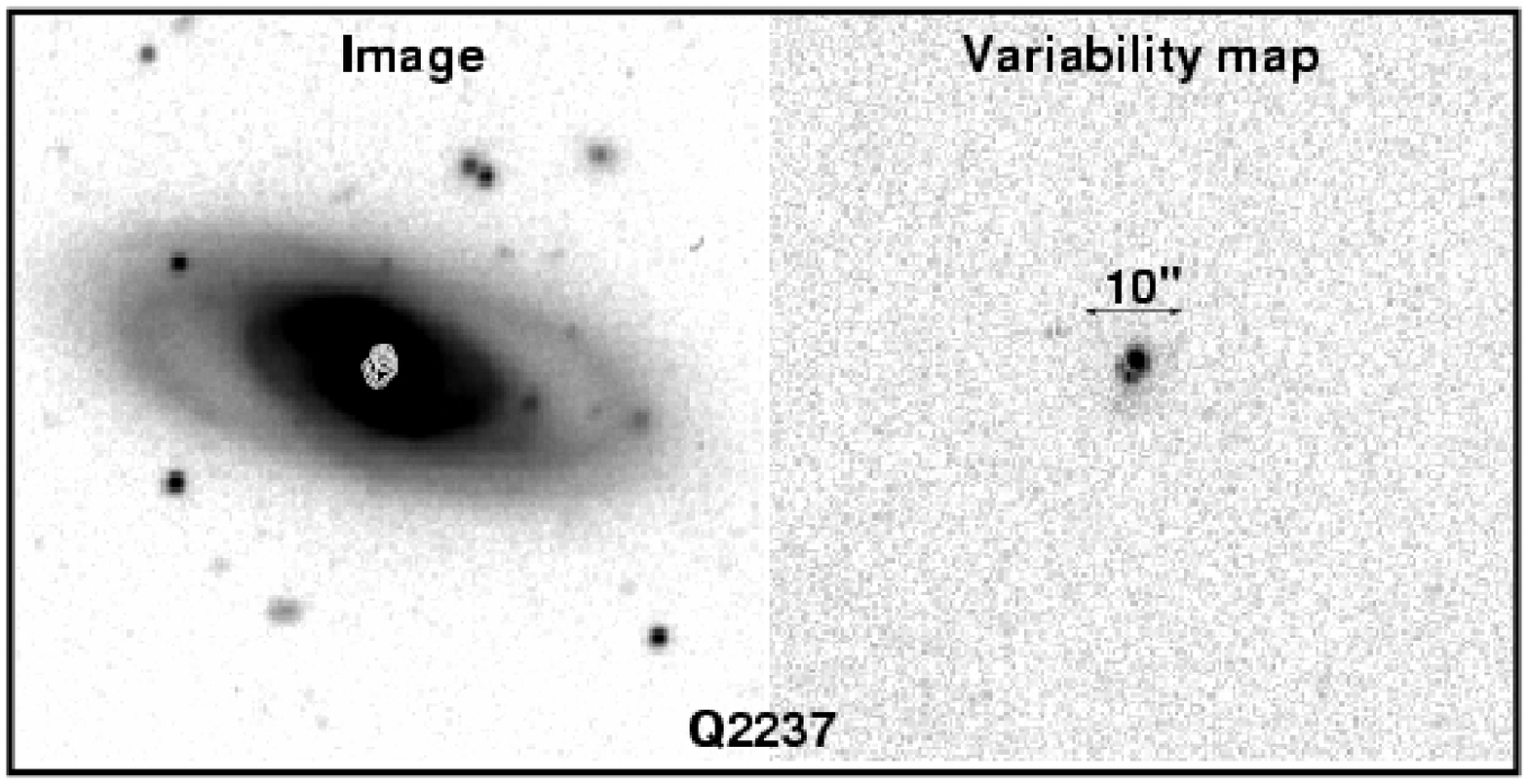}
\caption{ 
  Average ($I_T$, left) and variability ($R_R/I_N$, right) images of 
  RXJ~1131--1231 (top)  and Q~2237+0305 (bottom).
  The Einstein radii of the two lenses are 3\farcs2 and
  1\farcs8 respectively. 
  }
\label{fig:lenses2}
\end{figure}

\section{Demonstrations}
\label{sec:demonstrations}

In order to demonstrate our idea, we analyzed our monitoring data for four lenses using
the ISIS difference imaging package (Alard~\citeyear{Alard2000p363}).  Each epoch consisted
of a 15~min R-band exposure (taken as three 5~min exposures) with the SMARTS 1.3m telescope 
using ANDICAM (Depoy et al.~\citeyear{Depoy2003}).  We analyzed only the images with seeing FWHM $\leq 1\farcs5$, sky
backgrounds $\leq 1000$~ADU/pixel, and flux calibrations relative to the reference
image $\geq 0.5$ (to eliminate data taken through clouds).  The four lenses we
considered were 
QJ~0158--4325 (Morgan et al.~\citeyear{Morgan1999}), 
SDSS~0924+0219 (Inada et al.~\citeyear{Inada2003}) 
RXJ~1131--1231 (Sluse et al.~\citeyear{Sluse2003}), 
and Q~2237+0305 (Huchra et al.~\citeyear{Huchra1985}).   
RXJ1131--1231 consists of a bright cusp triple spanning 2\farcs3 separated by 
3\farcs2 from a much fainter fourth image.  SDSS~0924+0219 is
a more compact four-image lens. It has an Einstein ring diameter of 1\farcs7  
and the flux is dominated by the brightest image.   QJ~0158--4325 is a still more
compact two-image lens with an image separation of $1\farcs2$.  In our monitoring
program we only include lenses with image separations greater than $1\farcs0$, so
we cannot provide examples with smaller separations.  This is not a significant bias
since the median separation of gravitational lenses is approximately $1\farcs4$
(e.g. Browne et al.~\citeyear{Browne2003p13}) and the surveys should have
better resolution than our SMARTS data..  We do, however, include
Q2237+0305, a four image lens with a 1\farcs8 Einstein ring diameter buried in the
bulge of a very bright (B$\sim 15$~mag) low redshift spiral galaxy.  We used
71, 37, 57 and 26 epochs of data for QJ~0158--4325, RXJ~1131--1231, 
SDSS~0924+0219, and Q~2237+0305, respectively.
  
For each lens we computed the average of the images  $I_T =  (1/N) \sum_i I_i$, its estimated
noise $N_T$, and the rms of the difference images $R_R^2 = (1/N)\sum_i R_i^2$.  In 
computing $R_R$,  we rejected the two epochs with maximum values of $|R_i|$ for each pixel 
in order to remove satellite trails and low level cosmic ray events which had not been found 
by earlier processing procedures.  The overall signal-to-noise ratio in the combined images 
is very high, with a point source sensitivity of roughly R$\simeq 25$~mag. 
We compute the significance of the variability using the ratio $R_R/N_T$ between
the variance image and the noise in the average image, a ratio
which should be unity in the absence of variability or systematic errors.  In
practice, our rms images $R_R$ are limited by systematics beyond the point
(roughly 10 images) that the statistical errors approach 1\% of the sky level. 
The limitation is presumably due to systematic problems associated with flat 
fielding, interpolation, and the difference imaging algorithms. 

Figs.~\ref{fig:lenses1} and \ref{fig:lenses2} show our four examples, comparing the 
average image $I_T$ to a map of the significance of the variability, $R_R/N_T$.   
The first point to note is the complete vanishing of everything in the field 
other than the lens.  The second point to note is that all four lenses are
easily recognized as multi-component/extended sources without any need for
further processing.  The appearance of the four-image lenses (SDSS~0924+021,
RXJ~1131--1231 and Q~2237+0305) is particularly striking.  To be fair, this
is true for all these lenses but Q~2237+0305 in the direct images as well.
On the other hand, emission from the lens galaxy masks a steadily increasing fraction of lensed 
quasars fainter than 20~mag, so Q~2237+0305 is more ``typical'' of the 
faint quasar lenses that will comprise the majority of the systems
detectable in the deep synoptic surveys than the other three systems.  

\begin{figure}[t]
\plotone{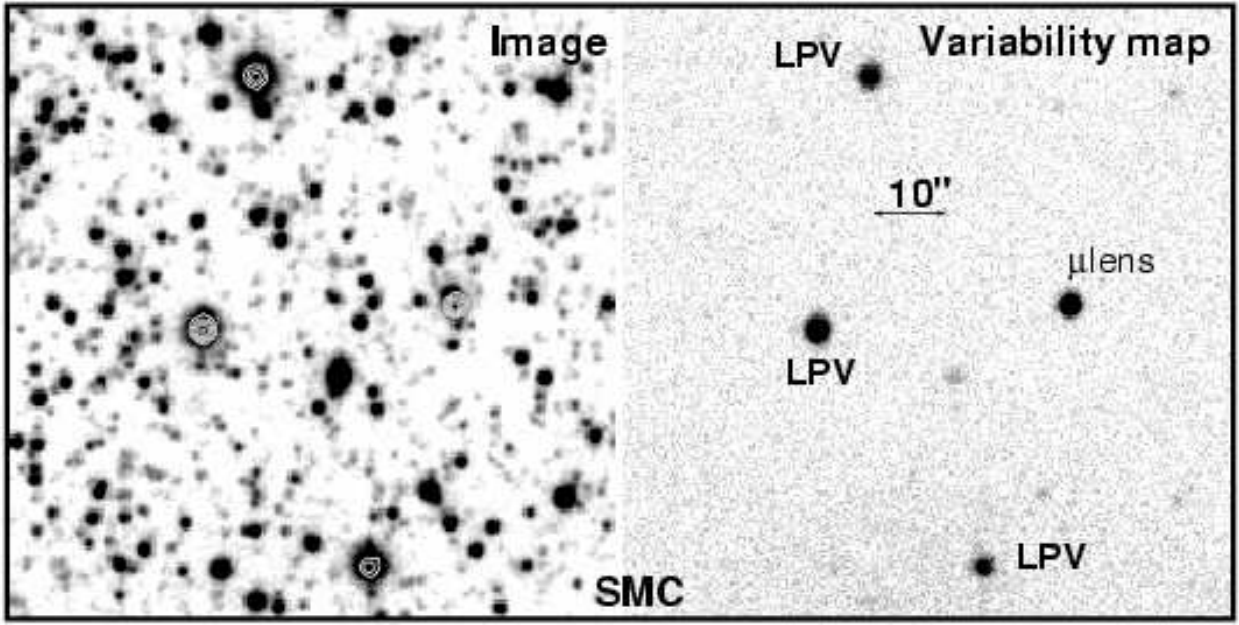}
\caption{
  Average ($I_T$, left) and variability ($R_R/I_N$, right) images of the field
  of the microlensing event OGLE-2005-SMC-001.  The microlensed source ($\mu$lens)
  and the three long period variable (LPV) stars in the field are labeled.
  Note that all four sources are stellar in the variability map even though
  they are blended with other stars in the direct image.
  }
\label{fig:smc}
\end{figure}

In the full 30~arcmin$^2$ fields, there were no other variable
sources. Thus, to provide a comparison to the behaviour of the
lenses, we applied the same procedures to the field of the microlensing event
OGLE-2005-SMC-001 using 62 5~min I-band observations obtained with 
ANDICAM by the MICROFUN collaboration (R. Pogge, private communication).
Fig.~\ref{fig:smc} shows the average and variability images for a region
with the same size as was used in Figs.~\ref{fig:lenses1} and \ref{fig:lenses2}.
In this case, the field contains four strongly variable sources, the
microlensing event and three long period variable (LPV) stars.  
All four variable sources are obviously 
stellar in the variability map and would not be flagged as lens
candidates even though they are all blended with other sources in the 
direct image.

\section{Summary}
\label{sec:summary}

Almost all new, large scale imaging surveys will be synoptic surveys that
monitor the time variability of sources in the survey area.  By using
difference imaging to search for extended variable sources, these surveys
can easily identify lensed quasars because almost all other variable
sources are either point sources or orbital tracks created by solar
system objects.  We estimate that gravitational lenses are the most
common extended variable sources for Galactic latitudes $|b| \gtorder 20^\circ$,
with modest contamination from pairs of variable stars, variable star/quasar
pairs and binary quasars.  Limiting the search to extended variable
sources reduces the number of non-lens background objects by more than
$10^5$. Thus, it should be relatively straight forward for SDSS, Pan-STARRS,
DES or LSST to identify the lensed quasars in their respective variability
survey areas.  For LSST, this should amount to roughly $10^3$ lensed
quasars to V$<23$~mag.

Note that the criterion of being an extended variable source can be combined
with other criteria to further reduce the rate of false positives based on 
other information available from the same survey.  For example, the quasars
will have slowly varying aperiodic light curves, while many stars will show
more rapid variability or periodic variability.  Where the source is resolved,
the light curves of lensed quasars should be similar and can be cross-correlated 
to measure a time delay and verify that the source is a lens.  Note, however,
that our discovery method depends only on the existence of variability
rather than the measureability of the delay.  The colors of 
stars and quasars are different at most redshifts (e.g. Richards et al.~\citeyear{Richards2004}), and the colors 
of lensed images should be similar, up to concerns about differential
extinction in the lens (e.g. Falco et al.~\citeyear{Falco1999}).   Finally, in the average image it should
be possible to detect a lens galaxy, potentially using difference imaging
methods to accurately subtract the quasar contribution.  In general,
the background of non-lens sources can be so greatly suppressed that we
suspect the only significant issue for candidate selection will be systematic
errors in identifying extended variable sources that are presently difficult
to quantify.

\acknowledgements 
We thank J. Hartman, B. Pindor and R. Scranton for discussions about aspects
of this paper.  We also thank R. Pogge and A. Gould for allowing us to use
the ANDICAM data for the OGLE-2005-SMC-001 event.

\end{document}